\documentstyle[preprint,aps]{revtex}
\tightenlines
\begin{document}

\title{Constraints on Brane-Localized Gravity}

\author{\normalsize{Philip D. Mannheim\footnote{Email address: 
mannheim@uconnvm.uconn.edu}} \\
\normalsize{Center for Theoretical Physics, Laboratory for Nuclear
Science and Department of Physics,
Massachusetts Institute of Technology, Cambridge, Massachusetts 02139} \\
\normalsize{and} \\
\normalsize{Department of Physics,
University of Connecticut, Storrs, CT 06269\footnote{permanent address}} \\
\normalsize{(MIT-CTP-2984, hep-th/0005226, September 8, 2000)} \\}

\maketitle

\begin{abstract}
In this paper we explore some general aspects of the embeddings associated 
with brane-localized gravity. In particular we show that the consistency 
of such embeddings can require (or impose) very specific relations between 
all the involved bulk and brane matter source parameters. We specifically
explore the embeddings of 3-branes with non-zero spatial 3-curvature $k$ into 
5-dimensional spacetime bulks, and show that for such embeddings, a
5-dimensional bulk cosmological constant is not able to produce the exponential
suppression of the geometry thought necessary to localize gravity to the brane.
\end{abstract}

\section{General Introductory Remarks}

It has been suggested recently 
\cite{Arkani-Hamed1998,Antoniadis1998,Randall1999a,Randall1999b} that it is
possible for our 4-dimensional universe to be a 3-brane\footnote{The 
dimensionality of a brane is defined by the number of its spatial dimensions
just like that of a sheet of material.} embedded in some
higher dimensional bulk spacetime whose spacelike extra dimensions need not in
fact be as  minuscule as their string theory Planck length expectation.
And while the original motivation of these studies was an attempt to solve the 
hierarchy problem, nonetheless the potential existence of any large such extra 
dimension is a matter of great interest in and of itself.\footnote{For a 
recent compilation of some of the rapidly growing literature in this field see 
e.g. \cite{Giddings2000}.} Moreover, Randall and Sundrum 
\cite{Randall1999a,Randall1999b} were able to show that the embedding of a 
flat Minkowski 3-brane into a 5-dimensional anti de Sitter
spacetime ($AdS_5$) would then  explicitly localize gravity to the
4-dimensional world, thereby releasing  the extra higher dimension from needing
to be tiny. Now while this is a very  nice property of the $AdS_5$ embedding,
it is important to see just how generic  it in fact is, and to what extent the
embedding into a 5-dimensional spacetime  of any given 4-dimensional set of
matter fields in any given 4-dimensional  geometry\footnote{In discussing 
embeddings it is important to recognize that the geometry of a given bulk can be
modified by the introduction into the bulk spacetime of a brane  with its own
specific symmetry, and with the brane setting up its own gravitational field in
the bulk, it is important to ascertain
whether a given $AdS_5$ bulk geometry remains so after the embedding
of the brane. To illustrate the point  we recall that the familiar
Schwarzschild de Sitter metric
$ds^2=-B(r)dt^2+dr^2/B(r)+r^2d\Omega$ (where
$B(r)=1-2MG/r-kr^2$) is a metric for which the geometry exterior to the source
is not maximally 4-symmetric ($R_{2323}=-kr^4(1+2MG/kr^3)sin^2\theta \neq
-k(g_{22}g_{33}-g^2_{23})$),  with it only becoming so asymptotically
far from the source. In the presence of the mass source then the
geometry exterior to the source is not the $dS_4$ geometry it would have been
had the source not been there.}  would in fact then  result in a gravity that
actually was localized  to the 4-dimensional world. Moreover, it is equally
necessary to see whether  any given such 4-space configuration of matter fields
can even be consistently  embedded into a higher dimensional space at all. And
indeed, in one sense the whole localization issue is initially somewhat
puzzling, since gravitating material always produces a gravitational field in
the empty space around it,  and it is straightforward to produce source
configurations in which the  associated gravitational potentials can actually
grow at distances far from such sources. While an immediately obvious example
of such a source might be a  uniform density, zero curvature sheet of
non-relativistic static gravitating material, a source  which produces a
constant, non-declining, Newtonian gravitational force away from the sheet when
the sheet is immersed in an otherwise flat,  empty background, such a
non-relativistic gravitational field is just a  coordinate artifact, being
equivalent to a uniform acceleration in flat space. Thus, as we will show
below, an embedding of such a sheet into a bulk with  a non-zero 5-dimensional
cosmological constant $\Lambda_5$ does in fact produce a gravity which is
localized to the sheet, though as we shall also see,  the covariantizing of
such a sheet (to then produce a true gravitational  field) proves to be
instructive, with the consistency of its embedding (even into a source free
bulk) being found to only be achievable for very  specific  brane equations of
state. Motivated by  this analysis, we shall then extend our study to the case
where the static  sheet or brane is endowed with a non-zero spatial 3-curvature
$k$ (a configuration whose embedding into a source free bulk leads to a
gravitational force which even grows with distance, one which is not a
coordinate artifact),  to then find in this case that the bulk $\Lambda_5$ is
found incapable of  producing exponentially suppressed localization of the
geometry to the brane.

To begin our analysis it is instructive to recall some general properties
of $AdS_5$ spacetime itself. As well as being constructible as a constant
surface  in a flat 6-dimensional space, the $AdS_5$ metric can also be given by
the  convenient form
\begin{equation}
ds^2=(R^2/z^2)(dz^2-dt^2+d\bar{x}^2)
\label{1}
\end{equation}
where $R$ is the radius of curvature. To see that this metric is in fact an 
$AdS_5$ metric, we note that since this metric is conformal to flat, we can 
explicitly determine its associated curvature by conformally transforming the 
flat $\eta_{\mu \nu}(x)$ metric according to $\eta_{\mu \nu}(x) \rightarrow 
\Omega ^2(z) \eta_{\mu \nu}(x)=g_{\mu \nu}(x)$ where $\Omega(z)=R/z$. Under 
such a transformation the initially zero 5-dimensional Ricci tensor is found 
to transform to
\begin{equation}
R^{\mu}_{\phantom{\mu}\nu} \rightarrow  
\Omega^{-5}\partial_{\rho}\partial^{\rho}
(\Omega^3)\delta^{\mu}_{\nu}/3-3\Omega^{-1}\partial^{\mu}\partial_{\nu}
(\Omega^{-1})=4\delta^{\mu}_{\nu}/R^2.
\label{2}
\end{equation}
Moreover, since the Weyl tensor vanishes in geometries which are conformal to 
flat, the Riemann tensor associated with the metric of Eq. (\ref{1}) is 
determinable from its associated Ricci tensor $R_{\mu \nu}=4g_{\mu\nu}/R^2$ 
alone, with it then 
immediately being found to take the form $R_{\lambda \rho \sigma \nu}
=-(g_{\sigma \rho}g_{\lambda \nu}-g_{\nu \rho}g_{\lambda \sigma})/R^2$. We thus 
recognize the spacetime associated with the metric of Eq. (\ref{1}) to be 
that of a 5-space with constant negative curvature $K=-1/R^2$, viz. $AdS_5$. 
As the above analysis shows, we could construct metrics of the form
$ds^2=(R^2/x^2)(dz^2-dt^2+d\bar{x}^2)$ where $x$ is any one of the four 
spacelike coordinates and still have an $AdS_5$ spacetime. However, because
of the signature change in the flat d'Alambertian operator 
$\partial_{\rho}\partial^{\rho}$, the metric 
$ds^2=(R^2/t^2)(dz^2-dt^2+d\bar{x}^2)$ would have constant positive curvature
$K=+1/R^2$ and thus be a de Sitter rather than an anti de Sitter space.
For negative curvature spaces then the multiplying overall factor in the
metric of Eq. (\ref{1}) must only be associated with one of the spacelike 
coordinates. Since for such coordinates the transformation $z=Re^{y/R}$ 
allows us to rewrite the metric in the form
\begin{equation}
ds^2=dy^2-e^{-2y/R}(dt^2-d\bar{x}^2)
\label{3}
\end{equation}
we see that in $AdS$ spaces the spatial exponential $e^{-2y/R}$ factor acts 
just like its temporal analog $e^{2t/R}$ in de Sitter spacetimes. Such 
exponential behavior in $AdS$ spaces is thus the spatial analog of inflation, 
with the $e^{-2y/R}$ factor leading to rapid suppression as we go out in $y$ 
away from the 4-dimensional space associated with the metric $dt^2-d\bar{x}^2$. 
Now since the multiplying factor in the metric of Eq. (\ref{1}) is quadratic
in $z$ (and uniquely so for $AdS$ spaces\footnote{Multiplying a flat space
metric by any conformal factor will always lead to a new metric which is 
conformal to flat. However, it is only the choice $\Omega^2(z)=R^2/z^2$ which 
leads to a spacetime with the same maximal number of Killing vectors as the 
original flat spacetime itself.}), we can make transformations of either of 
the forms $z=Re^{+y/R}$ and $z=Re^{-y/R}$ on the metric of Eq. (\ref{1}). Thus 
we can make
the transformation $z=Re^{+y/R}$ in the $y>0$ region and the transformation
$z=Re^{-y/R}$ in the $y<0$ region to then give us $e^{-2|y|/R}$ exponential 
suppression for every value of $y$, positive or negative. However, now the 
two regions in $y$ will be two separate patches of $AdS_5$ with there then 
necessarily being a discontinuity at $y=0$ where the two patches meet. It is 
thus at just such a discontinuity that our 4-dimensional universe can be 
located with our universe then being a 3-brane embedded in a higher dimensional 
bulk space containing two separate 
geometrical patches; with gravity potentially then being localized to the brane 
through the $e^{-2|y|/R}$ suppression 
\cite{Randall1999a,Randall1999b}.\footnote{The dependence of the geometry away 
from the brane on $|y|$ rather than on $y$ itself is characteristic of the 
requirement made in all brane localized gravity studies (this one included) 
that there is to be a 
$y \rightarrow -y$ symmetry of the metric around the $y=0$ brane.} It is  
thus to the implications of the embedding of such brane universes, and to the 
dynamical interplay of the bulk and the brane entailed by the very fact of 
such embeddings (even when the bulk itself contains no matter fields 
at all), to which we now turn, first in a 
source free empty background and then in one with a bulk $\Lambda_5$ (viz. a 
bulk that would be $AdS_5$ in the absence of the brane).

\section{Embedding a brane in an empty bulk}

For the embedding of a homogeneous, isotropic standard 4-dimensional 
Robertson-Walker universe with spatial 3-curvature $k$ into an arbitrary
(and thus not necessarily $AdS_5$) 5-dimensional bulk space, the most general
allowed maximally $r,\theta,\phi$ 3-symmetric metric takes the generic form
\begin{equation}
ds^2=-n^2(y,t)dt^2+a^2(y,t)[dr^2/(1-kr^2)+r^2d\Omega] +b^2(y,t)dy^2 
+2c(y,t)dtdy,
\label{4a}
\end{equation}
up to arbitrary coordinate transformations involving $y$ and $t$. Recognizing
the pure $y,t$ sector of this metric to be the most general 2-dimensional 
metric in a $y,t$ space, we can thus make a coordinate transformation in this
space to remove the $y,t$ cross-term, and in the illustrative static limit 
which we study here can then reabsorb $b(y)$ into a redefinition of $y$ to
thus yield 
\begin{equation}
ds^2=-n^2(y)dt^2+a^2(y)[dr^2/(1-kr^2)+r^2d\Omega] +dy^2 
\label{4}
\end{equation}
a metric whose embedding and localization of gravity aspects we now
study.\footnote{While we  concentrate here on embedding and localization 
issues, the brane theory associated with Eq. (\ref{4a}) has also been studied 
as a cosmology in and of itself, see e.g.
\cite{Csaki1999,Binetruy2000,Mohapatra2000}.} While in brane-localized studies
it is desired to recover standard gravity only in the 4-dimensional world, it
is conventional to assume that the full 5-space  gravity is given simply by the
5-dimensional Einstein equations (rather than  by some more complicated set of
5-dimensional equations), viz.
\begin{equation}
G_{AB}=R_{AB}-g_{AB}R^C_{\phantom{C}C}/2=-\kappa^2_5 [T_{AB} 
+T_{\mu \nu}\delta^{\mu}_A\delta^{\nu}_B\delta(y)]
\label{5}
\end{equation}
where $T_{AB}$ ($A,B=0,1,2,3,5$) is due to sources in the bulk and 
$T_{\mu \nu}$ ($\mu,\nu=0,1,2,3$) is due to sources on the $y=0$ brane.
For the symmetry of Eq. (\ref{4}) both of these energy-momentum tensors are 
given as perfect fluids, viz.
\begin{equation}
T^A_{\phantom{A}B}=diag(-\rho_B,P_B,P_B,P_B,P_T),~~ 
T^{\mu}_{\phantom{\mu}\nu}=diag(-\rho_b,p_b,p_b,p_b)
\label{6}
\end{equation}
($B$ denotes bulk and $b$ denotes brane). For the metric of Eq. (\ref{4}) it is
straightforward to write the 5-dimensional  Einstein equations, with the 
resulting expressions simplifying 
to (the prime denotes differentiation with respect to $y$) 
\begin{equation}
G_{00}=3e^2f^{\prime\prime}/2f^2-3e^2k/f^2=-\kappa^2_5 e^2[\rho_B
+\rho_b\delta(y)]/f
\label{7}
\end{equation}   
\begin{equation}
G_{ij}=[-f^{\prime\prime}/2-fe^{\prime\prime}/e+k]\gamma_{ij}
=-\kappa^2_5 f[p_B+p_b\delta(y)]\gamma_{ij}
\label{8}
\end{equation}
\begin{equation}
G_{55}=-3f^{\prime}e^{\prime}/2fe+3k/f=-\kappa^2_5 P_T
\label{9}
\end{equation}   
when the identification $a(y)=f^{1/2}(y)$, $n(y)=e(y)/f^{1/2}(y)$ is made 
[here $\gamma_{ij}=diag(1/(1-kr^2),r^2,r^2sin^2\theta)$]. 

While we shall discuss the implications of these equations in various 
situations below, we note immediately that when the bulk $T_{AB}$ is set to 
zero and when the spatial $i,j$ coordinates are restricted to a flat 
2-dimensional $(x,y)$ plane and $y$ is replaced by the usual spatial $z$, 
Eq. (\ref{4}) then describes a uniform, infinite, flat 2-dimensional sheet of 
static matter embedded in (what otherwise would have been) ordinary empty
spacetime, a system whose Newtonian limit is known to correspond to a
gravitational potential which grows linearly  with $z$ and a gravitational
force $F(z)$ per unit mass which is independent  of $z$. 
Moreover, in such a case, the Newtonian gravitational force points toward the 
$z=0$ sheet no matter which side we consider, with Gauss' Law yielding
$F(z=0^{+})-F(z=0^{-})=4\pi G\sigma$ and 
$F(z=0^{+})=-F(z=0^{-})=2 \pi G\sigma$
for a sheet of surface matter density $\sigma$, with the gravitational
potential $\phi=2\pi G\sigma |z|$ thus being discontinuous across the 
surface.\footnote{With brane studies always requiring the gravitational 
potential to only be a function of $|z|$, we see that such a requirement
nicely dovetails with the constraints of Gauss' Law.}  As such the
relation $F(z=0^{+})-F(z=0^{-})=4\pi G\sigma$
is a non-relativistic analog of the fully covariant relativistic Israel 
junction 
conditions \cite{Israel1966} (see e.g. \cite{Chamblin1999,Davidson1999} for 
some recent derivations)
\begin{equation}
K_{\mu\nu}(y=0^{+})-K_{\mu\nu}(y=0^{-})=-\kappa^2_5(T_{\mu\nu}-
q_{\mu\nu}T^{\alpha}_{\phantom{\alpha}\alpha}/3)
\label{10}
\end{equation}   
across a discontinuous surface with normal $n^{A}$ where $q_{AB}=
g_{AB}-n_An_B \equiv q_{\mu\nu}$ is the induced metric on the surface and
$K_{\mu\nu}=q^{\alpha}_{\phantom{\alpha}\mu}q^{\beta}_{\phantom{\beta}\nu}
n_{\beta ; \alpha}$ is its extrinsic curvature. We shall thus expect to see
an analog of this Newtonian gravity discontinuity in the treatment of the 
relativistic case associated with Eq. (\ref{4}), something to which we now 
turn.

For the simplest case first of a $k=0$ spatially flat Robertson-Walker 
geometry embedded in a source free bulk, on taking the metric
coefficient $a(y)$ to be a function of $|y|=y[\theta(y)-\theta(-y)]$ (where  
$|y|^{\prime}= \theta(y)-\theta(-y)$, $|y|^{\prime 2}=1$, 
$|y|^{\prime \prime}=2\delta(y)$), integration of Eq. (\ref{7}) is then 
found to yield 
\begin{equation}
a^2(y)=\alpha(1-\kappa^2_5\rho_b |y|/3)
\label{11}
\end{equation}   
where $\alpha$ is an arbitrary constant which can be absorbed in a redefinition
of the spatial $x_i$ coordinates. With Eq. (\ref{9}) then obliging 
$e(y)=e(|y|)$
to be a constant, we can then set $n(y)=1/a(y)$, with Eq. (\ref{8}) then
recovering the solution of Eq. (\ref{11}) provided
\begin{equation}
p_b=-\rho_b/3.
\label{12}
\end{equation}   
As a check on this solution, we note that the Israel junction conditions which
follow from Eq. (\ref{10}) in our particular case, viz. \cite{Binetruy2000} 
\begin{equation}
(a^{\prime}(y=0^{+})-a^{\prime}(y=0^{-}))/a(y=0)
=-\kappa^2_5\rho_b/3, 
\label{12a}
\end{equation}   
\begin{equation}
(n^{\prime}(y=0^{+})-n^{\prime}(y=0^{-}))/n(y=0)=
\kappa^2_5(3p_b+2\rho_b)/3, 
\label{12b}
\end{equation}   
are indeed satisfied by our obtained discontinuity.\footnote{With all of the 
metric coefficients being functions of $|y|$, the junction conditions require
the terms linear in $|y|$ to be first order in $\kappa^2_5$, viz.
$a(|y|)=\alpha(1-\kappa^2_5\rho_b |y|/6) +O(|y|^2)$,
$n(|y|)=\beta(1+\kappa^2_5(3p_b+2\rho_b) |y|/6) +O(|y|^2)
=\beta(1+\kappa^2_5\rho_b |y|/6) +O(|y|^2)$. The junction conditions thus 
generate a contribution to 
the bulk Riemann tensor (see Eq. (\ref{29}) below) which only begins in order 
$\kappa^4_5$, something which is to be expected since the order $\kappa^2_5$ 
constant Newtonian gravitational acceleration associated with a metric with 
$n(y)=\beta(1+\kappa^2_5\rho_b |y|/6)$ is removable by a coordinate 
transformation, with coordinate independent gravitational effects thus only
beginning in order $\kappa^4_5$, with true gravity thus needing the
$O(\kappa^2_5)$ terms of both of the 
$n(y)$ and $a(y)$ metric coefficients to be non-zero.} 
With 
$n^2(y) \rightarrow 1+\kappa^2_5\rho_b |y|/3$ in the weak gravity limit,
we see that we nicely recover the linear potential characteristic of a 
Newtonian gravity sheet (though, as will be clarified below, one actually 
not with the standard weak gravity coefficient), with gravity not at all being 
localized to the brane, 
and with the strong gravity limit even possessing a 
singularity at $|y|=3/\kappa^2_5\rho_b$.

While we thus see that we can obtain the anticipated non-localized solution, we 
find that it is only obtainable for a very particular equation of state, one 
with a negative pressure.\footnote{To see that this is in fact a generic effect 
we note that for a flat 2-brane embedded in an empty 4 space, viz. one 
described by the metric $ds^2=-n^2(z)dt^2+a^2(z)(dx^2+dy^2)+dz^2$, the 
4-dimensional Einstein equations $G_{\mu\nu}=-\kappa^2_4T_{\mu\nu}$ take the
form $G^0_{\phantom{0}0}=-(2aa^{\prime \prime}
+a^{\prime 2})/a^2=\kappa^2_4\rho_b\delta(z)$, $G^x_{\phantom{x}x}
=-(a^{\prime \prime}n+a^{\prime}n^{\prime}+an^{\prime \prime})/an=
-\kappa^2_4p_b\delta(z)$, $G^z_{\phantom{z}z}=-a^{\prime}(a^{\prime}n+
2an^{\prime})/a^2n=0$, with solution 
$a(z)=1/n^2(z)=(1-3\kappa^2_4\rho_b|z|/8)^{2/3}$, constraint 
$p_b=-\rho_b/4$, Israel junction conditions of the form 
$K_{\mu\nu}(y=0^{+})-K_{\mu\nu}(y=0^{-})=-\kappa^2_4(T_{\mu\nu}-
q_{\mu\nu}T^{\alpha}_{\phantom{\alpha}\alpha}/2)$ (viz.
$(a^{\prime}(y=0^{+})-a^{\prime}(y=0^{-}))/a(y=0) =-\kappa^2_4\rho_b/2$,
$(n^{\prime}(y=0^{+})-n^{\prime}(y=0^{-}))/n(y=0)=
\kappa^2_4(2p_b+\rho_b)/2$), and a Riemann tensor which is again of order
$\kappa^4_4$. In passing we note also that our result confirms an old result
of Vilenkin \cite{Vilenkin1983} that no static solution is possible for 
2-branes with equation of state $p_b=-\rho_b$. However, we now see that a 
static solution is possible when $p_b=-\rho_b/4$.} In order to understand this 
result we need to distinguish 
between the role that gravity plays in an ordinary 4-dimensional world and the
one that it appears to be playing in the embedded case. As regards 
first the 
conventional pure 4-dimensional situation with no embedding into a fifth 
dimension, we note that there the fluid equation of state is usually taken as
a fixed, gravity independent input, and the gravity which it produces is then
determined as output. Nonetheless, even in that case the $p_b/\rho_b$ ratio 
need 
not necessarily be positive. Thus even while the energy density and pressure 
of a high temperature ideal gas due to the kinematic 
motions of the gas particles are both positive, if the gas is cooled into a 
solid phase it then undergoes a
phase transition, a long range order effect such as condensation into an
ordered crystal lattice, an effect which can be associated with the negative
pressure characteristic of vacuum breaking,\footnote{In such a case the 
harmonic
phonon mode fluctuations in the lattice will still have positive pressure, it
is just that they have nothing to do with the mechanism which put the atoms
onto the lattice sites in the first place by minimizing the free energy.
Rather they are only a perturbation around such a minimum, a minimum to which
gravity however is sensitive.} with non-gravitational physics thus being 
capable of leading to fluids with negative pressure.\footnote{The 
$p_b=-\rho_b/3$ equation of state required above, for instance, 
could be associated 
with an isotropic network of cosmic
strings, with such a negative pressure fluid potentially leading to the cosmic
acceleration (see e.g. 
\cite{Mannheim1998} for a recent review) associated with quintessence models 
\cite{Caldwell1998}.} Thus, if the fluid 
equation of state is to be taken as a fixed, gravity independent input in
the embedded gravity case as well, we would have to conclude that unless the 
fluid actually possesses the needed equation of state, then no (static) 
embedding would 
in fact be 
possible.\footnote{In their original papers Randall and Sundrum noted the need 
to have a fixed relation between bulk and brane cosmological constants in 
models in which $n(y)$ was initially set equal to $a(y)$. We now see that 
restrictions on the structure of the energy-momentum tensor are of much broader 
generality, in principle involving the energy densities and pressures of all 
brane matter sources, restrictions which are intrinsic to all embeddings 
(even ones into source-free bulks) and not just only to those associated with
$AdS_5$.} 
   
However, in the brane-embedded case, it turns out that gravity can potentially 
play a different role, one in which it could be instrumental in actually fixing
the  fluid equation of state in the first place, with the equation of state
then  being output to the problem rather than input. In particular, the key
difference between the embedded and  non-embedded cases is that in the embedded
case the matter sources are  assumed to be confined to the brane, with no brane
matter contribution to $T_{55}$ being permitted. As a consequence, the $G_{55}$
component of the  empty bulk Einstein tensor has to vanish, a quite non-trivial
requirement  which has dynamical implications not present in a non
brane-embedded situation where the fluid is  otherwise free to flow in all
available spatial directions. Such a vanishing is then a constraint imposed by
the geometry,  and even when there are no explicit bulk matter fields to apply 
stresses on the brane, nonetheless there is still a non-vanishing Riemann
tensor in the bulk,\footnote{I.e. the very presence of the brane modifies the
geometry in the source-free bulk, to thus prevent it from being the flat one
that it would have been in the absence of the brane.} to thus enable the bulk
gravitational field to provide such stresses instead, with the bulk curvature
and the brane pressure then  potentially being able to dynamically adjust to
each other to thereby  fix the pressure on the brane and yield the brane
equation of state as  output.\footnote{Thus while the Israel junction condition 
$(n^{\prime}(y=0^{+})-n^{\prime}(y=0^{-}))/n(y=0)=
\kappa^2_4(2p_b+\rho_b)/2$  associated with the
embedding of a flat 2-brane in an empty 4 space would actually yield the 
standard weak gravity Gauss' Law 
$(n^{\prime}(y=0^{+})-n^{\prime}(y=0^{-}))/n(y=0)=
\kappa^2_4 \rho_b/2$ if we were to ignore the pressure $p_b$ on the brane (the
usual weak gravity assumption), we see that the consistency of the 
embedding requires a very 
different brane pressure, one of the same order of magnitude as $\rho_b$, to 
thus yield to a weak ($\kappa^2_4$ small) gravity limit whose potential 
has a different
normalization than that associated with a standard pressureless weak Newtonian 
gravity sheet.
(Since the standard Newtonian potential of a sheet is just a coordinate 
artifact, there is no reason for it to have to correspond to the 
non-relativistic limit of the true gravity associated with a fully 
covariantized 
uniform sheet.)} Thus rather than a given 
input fluid equation of state imposing an output geometry on gravity, gravity
instead could impose an output equation of state on the fluid.

The notion that the presence of an extra dimension might have dynamical 
implications and that higher dimensional gravity might play a role in 
stabilizing lower dimensional systems is certainly a very interesting one which 
requires further study. To illustrate its capability it is instructive to 
recall Einstein's attempt to construct a static 4-dimensional model of the 
universe. As is well-known, in his attempt to do so Einstein introduced a 
cosmological constant and was then able to find a non-trivial static 
universe solution provided 
the spatial 3-curvature $k$ of the universe was taken to be positive. In such a 
situation the ordinary 4-dimensional Einstein equations take the form 
$G^0_{\phantom{0}0}=3k=\kappa^2_4\rho_b$, $G^i_{\phantom{i}j}=k\delta^i_j=
-\kappa^2_4p_b\delta^i_j$ to precisely impose the selfsame $p_b=-\rho_b/3$ 
equation of state which we obtained above while fixing
$k=\kappa^2_4\rho_b/3$.\footnote{Einstein himself satisfied the relation 
$p_b=-\rho_b/3$ by taking $\rho_b=\rho_m+\lambda$, $p_b=-\lambda$ where
$\rho_m$ is the energy density of ordinary matter. In this solution then
ordinary matter was taken to have no pressure and the cosmological constant
was tuned to be given as $\lambda=\rho_m/2$. Thus already in this now quite 
ancient model we see the need for constraints (either input or output)
on the components 
of the 4-dimensional matter energy-momentum tensor. In passing we
note that current observations
\cite{Riess1998,Perlmutter1999,deBernardis2000} almost a  century later are
apparently requiring a similar such fine tuning  between 
4-dimensional matter and vacuum energy densities (for some remedies to this 
perplexing problem see e.g. \cite{Mannheim1999}).} As we now see, in the 
5-dimensional $k=0$ model discussed above, the binding role played by
positive $k$ in the 4-dimensional world is instead provided by the embedding, 
with the constraint 
condition $G_{55}=0$ then providing a dynamics not otherwise present 
in the 4-dimensional system itself.\footnote{Negative pressure solutions to the 
cosmic acceleration problem could thus arise as a 4-dimensional reflection of a 
higher dimensional embedding.} 

In fact this phenomenon is actually a quite 
general one. Specifically if the Einstein equations are assumed to hold in the 
bulk, it turns out \cite{Shiromizu1999,Sasaki1999} that the induced 
gravitational equations on the brane actually deviate from the standard 
4-dimensional 
Einstein equations, with the additional terms that are found being explicit 
consequences of the embedding, viz. they do not represent new matter sources 
in the 4 space, but rather they arise though the constraints associated with 
the very existence of the embedding. In particular the authors of  
\cite{Shiromizu1999} noted that since the difference between the 4-dimensional 
Riemann tensor $^{(4)}R^{\alpha}_{\phantom{\alpha} \beta \gamma \delta}$ of a
general 4-surface and the 5-dimensional Riemann tensor $R^A_{\phantom{A}BCD}$ of
some general 5-bulk into which it is embedded can be completely
characterized by  a function quadratic in the extrinsic  curvature tensor
$K_{\mu\nu}$ of the 4-surface according to the  Gauss embedding formula 
\begin{equation}
^{(4)}R^{\alpha}_{\phantom{\alpha} \beta \gamma \delta}=
R^A_{\phantom{A}BCD}q_A^{\phantom{A}\alpha}
q^B_{\phantom{B}\beta}q^C_{\phantom{C}\gamma}q^D_{\phantom{D}\delta}-
K^{\alpha}_{\phantom{\alpha} \gamma}K_{\beta \delta}+
K^{\alpha}_{\phantom{\alpha} \delta}K_{\beta \gamma},
\label{13}
\end{equation}   
use of the bulk Einstein equations, the Israel junction conditions at the
surface of the brane and the assumption of a $y\rightarrow -y$ symmetry around
the $y=0$ brane then enable us to express the 4-dimensional Einstein tensor in 
terms of quantities on the brane which must necessarily be quadratic in the 
energy-momentum tensor of the brane. In particular, for generic metrics of the 
form $ds^2=dy^2+q_{\mu\nu}dx^{\mu}dx^{\nu}$ and a brane energy-momentum 
tensor of the specific form\footnote{This particular form was chosen 
so that one of the quadratic terms would then yield a term linear in 
$\tau_{\mu \nu}$.}  
\begin{equation}
T^A_{\phantom{A}B}=-\Lambda_5\delta^A_{\phantom{A}B},~~
T_{\mu\nu}=-\lambda q_{\mu\nu}+\tau_{\mu\nu}
\label{14}
\end{equation}   
the authors of \cite{Shiromizu1999} found that the 4-dimensional Einstein 
tensor on the brane is given by
\begin{equation}
^{(4)}G_{\mu \nu}=\Lambda_4q_{\mu \nu}-8\pi G_{N}\tau_{\mu\nu}
-\kappa^4_5\pi_{\mu\nu}-\bar{E}_{\mu\nu}
\label{15}
\end{equation}   
where
\begin{equation}
G_{N}=\lambda\kappa^4_5/48\pi,~~
\Lambda_4=\kappa^2_5(\Lambda_5+\kappa^2_5\lambda^2/6)/2,~~
E_{\mu\nu}=C^A_{\phantom{A}BCD}n_{A}n^{C}q^B_{\phantom{B}\mu}
q^D_{\phantom{D}\nu}
\label{16}
\end{equation}   
\begin{equation}
\pi_{\mu\nu}=-\tau_{\mu\alpha}\tau_{\nu}^{\phantom{\nu}\alpha}/4
+\tau^{\alpha}_{\phantom{\alpha}\alpha}\tau_{\mu \nu}/12
+q_{\mu\nu}\tau_{\alpha\beta}\tau^{\alpha\beta}/8
-q_{\mu\nu}(\tau^{\alpha}_{\phantom{\alpha}\alpha})^2/24,
\label{17}
\end{equation}   
and where $\bar{E}=[E(y=0^{+})+E(y=0^{-})]/2$ is the mean value of $E_{\mu
\nu}$ at the brane. Thus, even in the event that
gravity gets to be localized to the brane, we see  in general that on the brane
we would expect gravity to depart from that  given by just the standard
4-dimensional Einstein equations associated with a  non-embedded 4-dimensional
world.\footnote{For a perfect fluid
$\tau_{\mu \nu} =(\rho_m+p_m)U_{\mu}U_{\nu}+p_mq_{\mu \nu}$ for instance, the
additional  
$\pi_{\mu \nu}$ tensor takes the form 
$\pi_{\mu \nu}=[U_{\mu}U_{\nu}(2\rho_m^2+2\rho_m p_m)+q_{\mu\nu}(\rho_m^2
+2\rho_m p_m)]/12$,
and thus acts like an additional perfect fluid with 
pressure $P=(\rho_m^2+2\rho_m p_m)/12$ and 
energy
density $R=\rho_m^2/12$ (so that if $p_m=-\rho_m/3$, $P=+R/3$).} Thus 
measurements within a 4-dimensional world embedded in a higher dimensional 
bulk would in principle be able to reveal the presence of the higher 
dimensional space even if the gravitational field is localized to the 
4-dimensional world.\footnote{Noting the special role played by the brane 
cosmological
constant $\lambda$ in establishing the Newton constant term in Eq. (\ref{15}),
we see that the effective Newton constant $G_N$ would vary (possibly even in 
sign as well as magnitude) in different epochs 
separated by phase transitions, with early universe 
cosmology then potentially no longer being controlled by the Newton constant 
measured in a low energy Cavendish experiment. 
It is thus of interest to note that it is precisely an epoch dependence 
to both the sign and magnitude of the effective gravitational coupling 
constant which has recently been 
identified \cite{Mannheim1999} as a possible solution to the cosmological 
constant problem.} \footnote{In passing, we also note a subtlety in applying 
Eq. (\ref{15}) to the Schwarzschild problem. Specifically, even though the
4-dimensional $R_{\mu\nu}=0$ vacuum Schwarzschild solution can
\cite{Brecher2000} explicitly be embedded into a 5-space with a bulk
cosmological constant according to the localizing 
$ds^2=e^{-2|y|}[dr^2/(1-2MG/r)+r^2d\Omega-(1-2MG/r)dt^2]+dy^2$, a case where
every single term in Eq. (\ref{15}) is found to vanish, nonetheless, the bulk
is not $AdS_5$ in this case. Specifically, Eq. (\ref{15}) only involves the
projections of the Weyl tensor along the normal direction to the brane and
these components indeed do vanish in the bulk in the solution of
\cite{Brecher2000}. However, explicit calculation shows that the other
components of the Weyl tensor (viz. the ones not involved in Eq. (\ref{15})) do
not in fact vanish (cf. $C_{0101}=2MGe^{-2|y|}/r^3$), so that (just like in our
earlier discussion of the 4-dimensional Schwarzschild de Sitter metric) we find
that in the presence of Schwarzschild on the brane the geometry  cannot be the
maximally symmetric $AdS_5$ in the bulk (though it does become so
asymptotically far from the brane). Now in their paper the authors of
\cite{Shiromizu1999} showed that when there is a spatially  inhomogeneous
matter distribution on the brane, Eq. (\ref{15}) then prevents  the exterior 
bulk geometry from being pure $AdS_5$. Since setting
$R_{\mu \nu}=0$ on the brane actually requires a delta function singularity at
$r=0$ (if  the $1/r$ term in $g_{00}$ is to have a non-zero coefficient),
Schwarzschild on the brane actually entails an inhomogeneous source on the
brane, to thus yield a non $AdS_5$ bulk.}

To study further implications of such embeddings we turn now to our second
soluble model, namely a static 3-brane with non-zero $k$ embedded in an empty
bulk. In the event of non-zero $k$ the most general empty bulk solution to Eq. 
(\ref{7}) is directly given as 
\begin{equation}
a^2(y)=\alpha(1-\kappa^2_5\rho_b |y|/3+k|y|^2/\alpha),
\label{18}
\end{equation}   
with Eq. (\ref{9}) then leading to 
\begin{equation}
n^2(y)=(-\kappa^2_5\rho_b\alpha/3+2k|y|)^2/
\alpha(1-\kappa^2_5\rho_b |y|/3+k|y|^2/\alpha),
\label{19}
\end{equation}   
and with Eq. (\ref{8}) then entailing the equation of state
\begin{equation}
p_b=-\rho_b/3-12k/\alpha \kappa^4_5\rho_b.
\label{20}
\end{equation}   
Thus we see that the consistency of the embedding (cf. the non-trivial 
vanishing of $G_{55}$) again imposes constraints on the brane equation of
state, with the metric away from the brane now growing quadratically with 
distance. It is thus to the issue of whether or not there is to be a 
quenching of this metric when a bulk 
cosmological constant is introduced to which we now turn. 

\section{Embedding a brane in a non-empty bulk}

In the event of there being a bulk cosmological constant 
$-\rho_B=p_B=P_T=-\Lambda_5$ the structure of the solutions to the
5-dimensional Einstein equations will depend on whether there is a spatial 
curvature $k$ on the brane. Thus on setting $k=0$ first and taking $a(y)$ to 
be a function only of $|y|$ just as before, Eq. (\ref{7}) is then found to 
lead to 
\begin{equation}
{3 \over 2}{d^2 f(|y|) \over d|y|^2}=-\kappa^2_5\Lambda_5 f(|y|)
\label{21}
\end{equation}   
\begin{equation}
[3{df(|y|) \over d|y|}+\kappa^2_5\rho_bf(|y|)]\delta(y)=0.
\label{21a}
\end{equation}   
According to Eq. (\ref{21}) (viz. a pure bulk Einstein 
equation which would hold even in the absence of any brane at $y=0$) and 
its counterpart which comes from Eq. (\ref{8}), the most general allowed metric
coefficient in the $k=0$ case is then found to have an unbounded exponential
dependence
\begin{equation}
a^2(y)=f(y)=\alpha e^{\nu|y|}+\beta e^{-\nu|y|},~~n(y)f^{1/2}(y)=e(y)=\alpha
e^{\nu|y|}-\beta e^{-\nu|y|},
\label{22}
\end{equation}   
on distance if $\Lambda_5$ is negative (viz. anti de Sitter),
with the square of the exponent $\nu$ being given by
\begin{equation}
\nu^2=-2\kappa^2_5\Lambda_5/3.
\label{23}
\end{equation}
Thus no matter what sign we take for $\nu$, we see that 
in and of itself a bulk cosmological constant does not automatically lead to 
exponential suppression away from the brane in this case - as we recall from Eq.
(\ref{1}),  the $AdS_5$ metric is quadratic in $R/z$ - with the most general
solution to Eq. (\ref{21}) in fact necessarily being  unbounded, with having a
$\Lambda_5\neq 0$ bulk in and of itself thus not being sufficient to guarantee 
brane-localization of gravity.\footnote{In the original Randall-Sundrum study 
the brane geometry was taken to maximally 4-symmetric (viz. Minkowski), to thus
oblige the metric coefficients $a^2(y)$ and $n^2(y)$ to be equal to each other,
and thereby only allow solutions with a single exponential. (Whether this
single exponential itself would then be converging or diverging depends on
the sign of $\Lambda_b$.) However, once the brane geometry is lowered to
maximally 3-symmetric, two exponential terms with opposite sign exponents are
then allowed.} Thus we see that while the bulk might have been a pure $AdS_5$ 
bulk with only a single exponential in its metric in the absence of the 
brane, the introduction of the lower symmetry brane then lowers the symmetry in
the bulk with the unbounded exponential no longer automatically being
excluded.   

In order to try to remove this undesired growing exponential anyway, we note 
that the insertion of Eq. (\ref{22}) into the brane discontinuity 
formulas associated with Eq. (\ref{7}) and its Eq. (\ref{8}) counterpart 
yields
\begin{equation}
3\nu(\alpha-\beta)=-(\alpha+\beta)\kappa^2_5\rho_b,
\label{23d}
\end{equation}   
\begin{equation}
6\nu(\alpha+\beta)=(\alpha-\beta)\kappa^2_5(\rho_b+3p_b),
\label{23e}
\end{equation}   
relations whose solubility requires the quantity $\rho_b(\rho_b+3p_b)$ to
expressly be negative. Thus if in addition to the  equation of state (once again
we see that embeddings entail constraints on both the  bulk and brane matter
fields)
\begin{equation}
\Lambda_5-\kappa^2_5\rho_b(\rho_b+3p_b)/12=0
\label{23b}
\end{equation}   
which then follows, we additionally now impose the condition
\begin{equation}
p_b=-\rho_b
\label{25}
\end{equation}   
we will then explicitly force the coefficient $\beta$ to vanish and
reduce the metric to just one exponential term. Thus on dropping the $\beta$
dependent  term in Eq. (\ref{22}) (a point we shall return to  below), and
retaining only the $\alpha$ dependent one, we then find that  
the  discontinuity condition at the brane then fixes the sign of $\nu$
according
\begin{equation}
\nu=-\kappa^2_5\rho_b/3,
\label{22a}
\end{equation}   
with the geometry now being localized to the brane according to 
\begin{equation}
a^2(y)=\alpha e^{\nu|y|},~~n^2(y)=\alpha
e^{\nu|y|},
\label{22g}
\end{equation}   
when $\rho_b$ is 
positive (for $\rho_b$ negative no localization would be
obtained),  with Eq. (\ref{23b}) then yielding the 
compatibility condition
\begin{equation}
\Lambda_5+\kappa^2_5\rho_b^2/6=0.
\label{22b}
\end{equation}   
Thus when $\beta$ is set to zero the exponential dependence associated
with  Eq. (\ref{22g}) nicely quenches the linear metric
dependence found in Eq. (\ref{11}) in the 
empty bulk case just as desired, while precisely imposing on the brane the 
equation of state associated with a cosmological constant $\lambda$, with the
condition $-\Lambda_5=\kappa^2_5\rho_b^2/6\equiv \kappa^2_5\lambda^2/6$ then 
entailing the vanishing of the net effective brane cosmological constant 
$\Lambda_4$ of the general Eq. (\ref{16}),\footnote{Equation (\ref{16}) thus 
explains why this condition is in fact quadratic in $\lambda$.} just as found 
in the original Randall Sundrum study. 

Returning now to the more general $\beta$ dependent case
given in Eq. (\ref{22}), we see that it is its more general equation of state 
$\Lambda_5-\kappa^2_5\rho_b(\rho_b+3p_b)/12=0$, rather than the 
restricted one of Eq. (\ref{22b}), which actually matches on continuously to the
$p_b=-\rho_b/3$ equation of state obtained earlier as Eq. (\ref{12}) in the
$\Lambda_5=0$ case. The reason for this is due to the limiting process needed
to extract the  term linear in $|y|$ as needed for Eq. (\ref{11}) from a
function only  containing exponentials, with it being only a linear combination
of two exponentials with appropriately chosen singular coefficients 
($\simeq (1\pm 1/\nu)$) which can generate a non-vanishing linear term when the 
exponent $\nu=(-2\kappa^2_5\Lambda_5/3)^{1/2}$ is allowed to go to zero. Hence,
by not retaining the $\beta$ dependent term, we have then taken a $\Lambda_5$
dependent metric, viz. that of Eq. (\ref{22g}), whose $\Lambda_5 \rightarrow 0$ 
limit does not generate any term linear in $|y|$. It is thus our assumed 
$\beta$ vanishing $|y|=\infty$  boundary 
condition which takes care of the Eq. (\ref{11}) empty bulk linear term, with
the parameter $\beta$ itself actually only vanishing when  the very specific
equation of state
$p_b=-\rho_b$ is imposed,\footnote{Since we should in principle be able to
consider brane field configurations other than this  particular one, we see
that in general some configurations lead to suppression and some do not.} to
thus then lead to a geometry which is exponentially suppressed as we go away
from the brane. As we shall now show, however, in the presence of a non-zero
spatial curvature, i.e. in the presence of a non-trivial topology on the brane,
even picking a brane field configuration which  retains only the exponentially
damped
$\alpha$ dependent term will not in fact prove sufficient to  suppress the
geometry away from the brane.
  
In the $k\neq 0$ non-empty bulk case, Eqs. (\ref{22}), (\ref{23d}) and
(\ref{23e}) are found to be replaced by  
\begin{equation}
f(y)=\alpha e^{\nu|y|}+\beta
e^{-\nu|y|}-2k/\nu^2,~~e(y)=\alpha e^{\nu|y|}-\beta e^{-\nu|y|},
\label{25a}
\end{equation}   
\begin{equation}
3\nu(\alpha-\beta)=-(\alpha+\beta-2k/\nu^2)\kappa^2_5\rho_b,
\label{25b}
\end{equation}   
\begin{equation}
6\nu(\alpha+\beta)=(\alpha-\beta)\kappa^2_5(\rho_b+3p_b),
\label{25c}
\end{equation}   
where again $\nu^2=-2\kappa^2_5\Lambda_5/3$. If we now drop the $\beta e^{\nu
|y|}$ type term by hand by requiring the matter fields to obey 
$\kappa^2_5(3p_b+\rho_b)^2+24\Lambda_5=0$, the solution reduces to 
\begin{equation}
a^2(y)=\alpha e^{\nu|y|}+3k/\kappa^2_5\Lambda_5,~~n^2(y)=\alpha
e^{\nu|y|}/(1+3k e^{-\nu|y|}/\alpha\kappa^2_5\Lambda_5),
\label{26}
\end{equation}   
where
\begin{equation}
\nu=(-2\kappa^2_5\Lambda_5/3)^{1/2}=-\kappa^2_5\rho_b(1+
3k/\alpha\kappa^2_5\Lambda_5)/3,~~p_b=-\rho_b(1+2k/\alpha\kappa^2_5\Lambda_5),
\label{26a}
\end{equation}   
with the net brane cosmological constant $\Lambda_4$ vanishing this time 
if the brane matter energy density and brane cosmological constant (defined via 
$\rho_b=\rho_m+\lambda$) are fine-tuned according to
\begin{equation}
\lambda= -\rho_m(1+\alpha\kappa^2_5\Lambda_5/3k),
\label{26b}
\end{equation}   
so that the brane matter pressure defined via $p_b=p_m-\lambda$ is then 
related to the brane matter energy density according to $p_m=-\rho_m/3$.
As we thus see, even though the $AdS_5$ exponential damping factor does make
an appearance in the $k\neq 0$ case, nonetheless we find that $a^2(y)$ is not
asymptotically suppressed far away from the brane. Rather 
$a^2(y)$ tends to the non-vanishing value $3k/\kappa^2_5\Lambda_5$ if $k$ is
negative (even as $n^2(y)$ is then being suppressed), while if $k$ is positive 
$n(y)$ actually becomes singular. This lack of suppression is also evidenced
in the Riemann tensor, with its $R^{12}_{\phantom{12}12}$ component for 
instance, viz.
\begin{equation}
R^{12}_{\phantom{12}12}=f^{\prime 2}/4f^2 -k/f,
\label{29}
\end{equation}   
tending to $-\kappa^2_5\Lambda_5/3$ at large $y$ (viz. twice the pure $AdS_5$ 
value), with the bulk embedding thus never being able to 
counteract the effect of the spatial curvature of the brane, no matter how
judicious a choice of matter fields we may make. A similar situation is also
found for the Weyl tensor where\footnote{For metrics of the  form given in Eq.
(\ref{4}) all non-vanishing components of the Weyl tensor  are kinematically
proportional to
$C^{12}_{\phantom{12}12}$.} 
\begin{equation}
C^{12}_{\phantom{12}12}=[-2eff^{\prime \prime}+3ef^{\prime 2}-2efk-
3e^{\prime}ff^{\prime}+2f^2e^{\prime \prime}]/12ef^2
\label{30}
\end{equation}   
asymptotes to the non-vanishing value $-\kappa^2_5\Lambda_5/6$. Thus, unlike 
pure $AdS_5$, the metric 
associated with Eq. (\ref{26}) is not 
conformal to flat,\footnote{As Eq. (\ref{30}) also shows, even when 
$e(y)=f(y)=1$, $C^{12}_{\phantom{12}12}$ is equal to $-k/6$ and still does not 
vanish. Thus even while a constant curvature 3 space embedded in an otherwise 
flat 4 space (viz. standard 4-dimensional Robertson-Walker) produces a 
4-dimensional metric
which is conformal to flat, the same is not true of the same 3 space embedded 
in an otherwise flat 5 space, something a bulk cosmological constant is 
simply unable to alter. With only the $k=0$ Robertson-Walker geometry thus  
being localizable by an $AdS_5$ embedding, it would be of interest to see 
whether geometries with non-zero $k$ but with a negligibly small current era 
value of $\Omega_k(t)=-kc^2/\dot{R}^2(t)$ could still admit of an effective  
brane-localized current era gravity.} \footnote{It is possible to force the 
bulk Weyl tensor to vanish, to then force the bulk to actually be $AdS_5$.
However, this requires the reinstatement of the divergent $\beta$ type term,
with the coefficients in the metric of Eq. (\ref{25a}) having to then obey
$\alpha
\beta=k^2/\nu^4 \neq 0$, and with the fields then having to be related according
to
$\kappa^2_5\rho_b(2\rho_b+3p_b)-6\Lambda_5=0$. Thus even in this case  there is
still no localization of gravity. Further details of this particular case are
presented separately in
\cite{Mannheim2000}.} with the  bulk cosmological constant $\Lambda_5$ term 
not being able to  completely quench the quadratic growth previously found for
$a^2(y)$ in Eq.  (\ref{18}) in the $k \neq 0$, $\Lambda_5=0$
case.\footnote{While, as had  already been noted above, the bulk geometry would
not be pure $AdS_5$ in the  event that the brane matter source was spatially
inhomogeneous, we also see that even when the brane distribution is
homogeneous, the bulk geometry may still not be pure $AdS_5$.} Thus, to
conclude, we  see that while  embedding in a higher dimensional
$\Lambda_5<0$ bulk might lead  to a brane-localized geometry in certain
specific cases, it would appear  from study of our somewhat idealized static
cosmological model that such embeddings may not always lead to exponential
suppression  in general, with localization thus needing to be checked
on a case by case basis.

The author would like to thank Dr. A. Nayeri for introducing him to the subject 
of brane-localized gravity and would like to thank him and  Drs. A. H. Guth, 
H. A. Chamblin, D. Z. Freedman and A. Karch for some very helpful discussions.
The author  would also like to 
thank Drs. R. L. Jaffe and A. H. Guth for the kind hospitality of the Center 
for Theoretical Physics at the Massachusetts Institute of Technology where this 
work was performed. This work has been supported in part by funds provided by 
the U.S. Department of Energy (D.O.E.) under cooperative research agreement 
\#DF-FC02-94ER40818 and in part by grant \#DE-FG02-92ER40716.00.


\begin{thebibliography}{99}

\bibitem{Arkani-Hamed1998} N. Arkani-Hamed, S. Dimopoulos and G. Dvali, 
Phys. Lett. B {\bf 429}, 263 (1998).

\bibitem{Antoniadis1998} I. Antoniadis, N. Arkani-Hamed, S. Dimopoulos and G.
Dvali, Phys. Lett. B {\bf 436}, 257 (1998).  

\bibitem{Randall1999a} L. Randall and R. Sundrum, 
Phys. Rev. Lett. {\bf 83}, 3370 (1999).

\bibitem{Randall1999b} L. Randall and R. Sundrum, 
Phys. Rev. Lett. {\bf 83}, 4690 (1999).

\bibitem{Giddings2000} S. B. Giddings, E. Katz and L. Randall, JHEP 
0003:023 (2000). 

\bibitem{Csaki1999} C. Csaki, M. Graesser, C. Kolda and J. Terning, Phys. 
Lett. B{\bf 462}, 34 (1999).

\bibitem{Binetruy2000} P. Binetruy, C. Deffayet and D. Langlois, 
Nucl. Phys. B {\bf 565}, 269 (2000).

\bibitem{Mohapatra2000} R. N. Mohapatra, A. Perez-Lorenzana and C. A. de S. 
Pires, hep-ph/0003328.

\bibitem{Israel1966} W. Israel, Nouvo Cim. B {\bf 44}, 1 (1966). Erratum,
Nouvo Cim. B {\bf 48}, 463 (1967).

\bibitem{Chamblin1999} H. A. Chamblin and H. S. Reall, Nucl. Phys. B 
{\bf 562}, 133 (1999).

\bibitem{Davidson1999} A. Davidson and D. Karasik, Phys. Rev. D{\bf 60}, 
045002 (1999).

\bibitem{Vilenkin1983} A. Vilenkin, Phys. Lett. B {\bf 133}, 177 (1983).

\bibitem{Mannheim1998} P. D. Mannheim, Phys. Rev. D {\bf 58}, 103511 (1998). 

\bibitem{Caldwell1998} R. R. Caldwell, R. Dave and P. J. Steinhardt, Phys. Rev.
Lett. {\bf 80}, 1582 (1998).

\bibitem{Riess1998} A. G. Riess et. al., Astronom. J. {\bf 116}, 1009 
(1998).

\bibitem{Perlmutter1999} S. Perlmutter et. al., Astrophys. J. {\bf 517}, 
565 (1999).

\bibitem{deBernardis2000} P. de Bernardis et. al., Nature {\bf 404}, 
955 (2000).

\bibitem{Mannheim1999} P. D. Mannheim, astro-ph/9910093. 

\bibitem{Shiromizu1999} T. Shiromizu, K. Maeda and M. Sasaki,
Phys. Rev. D{\bf 62}, 024012 (2000).

\bibitem{Sasaki1999} M. Sasaki, T. Shiromizu and K. Maeda,
Phys. Rev. D{\bf 62}, 024008 (2000).

\bibitem{Brecher2000} D. Brecher and M. J. Perry, Nucl. Phys. B {\bf 566} 
151 (2000). 

\bibitem{Mannheim2000} P. D. Mannheim, MIT-CTP-2984; hep-th, September 8, 2000.



\end{thebibliography}
\end{document}